# More than magnetic isolation: Dynabeads as strong Raman reporters towards simultaneous capture and identification of targets

Jongwan Lee[1*†], Marissa McDonald[2*†], Nikiwe Mhlanga[1], Jeon Woong Kang[3], Rohit Karnik[1], and Loza F. Tadesse[1,4,5]*

[1]Department of Mechanical Engineering, MIT, Cambridge, MA, 02139, United States.
[2]Department of Health Sciences & Technology, MIT, Cambridge, MA, 02139, United States.
[3]Laser Biomedical Research Center, G. R. Harrison Spectroscopy Laboratory, MIT, Cambridge, MA, 02139, United States.
[4]Ragon Institute of Massachusetts General Hospital, MIT and Harvard, Cambridge, MA, 02139, United States.
[5]Jameel Clinic for AI & Healthcare, MIT, Cambridge, MA, 02139, United States.

[†]Indicates equal contribution
*Indicates to whom correspondence should be addressed;
E-mail: jongwanl@mit.edu, mmcdon33@mit.edu, lozat@mit.edu

**Abstract**

Dynabeads are superparamagnetic particles used for immunomagnetic purification of cells and biomolecules. Post-capture, however, target identification relies on tedious culturing, fluorescence staining and/or target amplification. Raman spectroscopy presents a rapid detection alternative, but current implementations target cells themselves with weak Raman signals. We present antibody-coated Dynabeads as strong Raman reporter labels whose effect can be considered a Raman parallel of immunofluorescent probes. Recent developments in techniques for separating target-bound Dynabeads from unbound Dynabeads makes such an implementation feasible with high specificity. We deploy Dynabeads anti-*Salmonella* to bind and identify *Salmonella enterica,* a major foodborne pathogen. Dynabeads present major peaks around 1000 and 1600 cm$^{-1}$ from aliphatic and aromatic C-C stretching of the polystyrene coating and near 1350 cm$^{-1}$ from the ɣ-$Fe_2O_3$ and $Fe_3O_4$ core, confirmed with electron dispersive X-ray (EDX) imaging. Minor to no contributions are made from the surface antibodies themselves as confirmed by Raman analysis of surface-activated, antibody-free beads. Dynabeads' Raman signature can be measured in dry and liquid samples even at single shot ~30 x 30 *μ*m area imaging using 0.5 s, 7 mW laser acquisition with single and clustered beads providing a 44- and 68-fold larger Raman intensity compared to signature from cells. Higher polystyrene and iron oxide content in clusters yields larger signal intensity and conjugation to bacteria strengthens clustering as a bacterium can bind to more than one bead as observed via transmission electron microscopy (TEM). Our findings shed light on the intrinsic Raman reporter nature of Dynabeads. When combined with emerging techniques for the separation of target-bound Dynabeads from unbound Dynabeads such as using centrifugation through a density media bi-layer, they have potential to demonstrate their dual function for target isolation and detection without tedious staining steps or unique plasmonic substrate engineering, advancing their applications in heterogeneous samples like food, water, and blood.

**Introduction**

Immunomagnetic separation is a technique that utilizes antibody-coated superparamagnetic beads that target antigens on cell surfaces to capture and concentrate cells. Since their invention more than 4 decades ago[1,2], immunomagnetic Dynabeads have been incorporated into routine biological experiments and even notable clinical trials such as in the isolation of CD34+ bone marrow-derived stem cells and CD3+/CD28+ T-cells in novel adoptive immunotherapy.[3] Dynabeads are the most frequently cited tool for immunomagnetic separation and result in high purity (95-100%) and viability (60-95%) of captured cells.[4,5] Particularly, their extremely versatile target-specific antibody-coupled surface enables the capture and magnetic separation of intact target cells from heterogeneous liquid samples such as blood and wastewater, eliminating the need for column separation or centrifugation techniques. Their superparamagnetic $\gamma$-$Fe_2O_3$ and $Fe_3O_4$ (hereby iron oxide) core and tunable surface functionalities enable them to be magnetically activated in the presence of an external magnetic field, allowing for the rapid and gentle isolation of target cells when needed and dispersion in solution when the magnetic field is no longer applied. Moreover, the polystyrene coating shields targets from the cytotoxic iron oxide core, making the beads biocompatible. Although cell sorting methods like fluorescence-activated cell sorting (FACS) are extremely effective, they require large concentrations of cells, tedious sample preparation steps, expensive cell-specific labels, and highly skilled personnel for use and maintenance, making them challenging for high throughput and field deployable applications.[6] Despite the immense benefits of Dynabeads in cell separation, detection of captured targets remains a challenge. Current approaches still rely on traditional, time consuming culturing or involved and tedious, molecular-based techniques such as polymerase chain reaction (PCR), enzyme immunoassays (EIA), and matrix-assisted laser desorption/ionization time of flight mass spectroscopy (MALDI-TOF MS) which suffer from downsides similar to FACS.[7–9] However, with emerging methods to separate free Dynabeads from Dynabeads bound to targets, such as by centrifugation through a density media bi-layer, specific detection of Dynabeads bound to targets in lieu of detecting the target pathogen itself may be possible.[10] Thus, a scalable and simple detection scheme leveraging this approach is needed to exploit the full versatility of Dynabeads for widespread use.

Raman spectroscopy is an emerging biosensing approach that can match the speed and versatility of Dynabeads for deployable applications. It generates molecular fingerprints of targets by using the inelastic scattering of light from samples.[11] In addition to scientific grade tabletop versions, cost effective, portable Raman systems are improving its accessibility for field applications. Thus, a Dynabead-based Raman spectroscopy assay presents an opportunity for rapid identification of magnetically concentrated targets. Studies in this area thus far, however, have focused on recording the signature of target cells or biomolecules, which typically generate weak Raman signals, and therefore relied on engineering complex dual substrates with a magnetic core and surrounding plasmonic metal nanoparticles such as gold and silver.[12–14] Engineering these unique substrates adds complex chemical synthesis schemes that can be challenging for widespread translation and are vulnerable to irreproducible enhancements of target signals, artifacts from antibodies on beads, and other contaminants in the sample due to off-target enhancement by metal nanoparticles. In contrast, similar to fluorescent probes, Raman probes can serve as a reporter of bound targets providing a strong Raman signature of their own fingerprint which can indirectly signal the presence of the target. However, adding such probes to substrates also suffers from similar drawbacks as plasmonic substrate engineering.

Here we report the use of magnetic Dynabeads as strong Raman reporters themselves without additional chemical labeling for simultaneous isolation and indirect detection of targets. We demonstrate detection of the Raman reporter signature of Dynabeads both in dried and liquid sample preparation formats. Specifically, we utilize Dynabeads coated with anti-*Salmonella* antibodies for indirect detection of *Salmonella enterica,* the leading cause of hospitalizations and death due to foodborne illness.[15,16] We note the dual function of Dynabeads coated with species-specific antibodies to both magnetically isolate cells in the sample and provide strong Raman signal towards a rapid, sensitive, and specific approach for indirect bacterial detection. This technique can be adapted to probe for any target using commercially available, pathogen specific Dynabeads in a lateral flow assay or vertical separation format to isolate pathogen bound-Dynabeads from free Dynabeads such as centrifugation with a density media bi-layer, granting specificity to this biosensing technique.[10] With these separation techniques, in combination with the Raman reporter property of Dynabeads demonstrated in this work, even a single to few pathogen bound-Dynabeads can present detectable Raman signature, providing indirect confirmation of the presence of a handful of pathogenic cells without the need for further culturing or staining steps.

**Results**

Our experimental setup for detection of *Salmonella* from liquid samples is summarized in Figure 1 below. Conjugates of Dynabeads anti-*Salmonella* and *S. enterica* were formed and suspended in a liquid well of deionized water (DIW) with optically transparent quartz substrate. Here, we aim to demonstrate the Raman reporter nature of Dynabeads in different sample types and measurement conditions, however the implementation of this Raman-based detection scheme is adoptable to the respective target-bound bead separation assay of choice to enable specificity. Taking advantage of the superparamagnetic property of Dynabeads, the conjugates were concentrated down to the imaging surface using a magnet immediately before interrogation with a 785 nm incident laser focused at the bottom of the well as shown in Figure 1A. This magnetic concentration step semi-fixed cell-bound beads in the field of view rather than having them freely float in the liquid. The target *S. enterica,* as shown in its transmission electron micrograph (TEM), is a rod-shaped bacteria with a 2-3 µm length and 0.7-1 µm width (Figure 1B). Dynabeads are ~2.5 µm diameter spherical particles with a polystyrene polymer shell covalently coupled to species-specific antibodies. As shown on Figure 1B and Figure S1, conjugation of Dynabeads ($10^7$ beads/mL) with *S. enterica* ($5 \times 10^6$ cells/mL) results in a tight, multi-cell interaction, demonstrating the effectiveness of the anti-*Salmonella* antibody coating to bind to *S. enterica* and support its isolation. The optical absorbance properties of Dynabeads anti-*Salmonella* and *S. enterica*-bound beads are shown in Figure 1C. The beads alone have a broad absorption peak at 500 nm with a smaller peak at 250 nm, matching peaks reported in other studies.[17] This absorption is primarily due to the iron oxide core, confirmed through further imaging and analysis discussed in Figure 5. *S. enterica* has high absorption in the UVC region (100-280 nm) but decreases steadily with increased wavelength; the high absorption at lower wavelengths is primarily due to nucleic acids in the bacteria.[18] Interestingly, the absorption of Dynabeads dampens following conjugation to *S. enterica.* This damping effect is likely due to increased light scattering from bound bacteria resulting in a lower detected absorption coefficient.[19,20] These absorbance trends are consistent across batches as shown in Figure S2. Furthermore, we note conjugation of beads to cells consistently maintains the clustering of Dynabeads in both liquid and dry sample preparation as shown in Figures 1B and S3. As shown in Figure 1B a single bacteria can bind to two beads at a time which strengthens and potentially promotes agglomeration upon conjugation of beads with bacteria, in cases where the number densities of the Dynabeads and bacteria in the solution are on the same order of magnitude. Details on sample preparation are included in Figure S4.

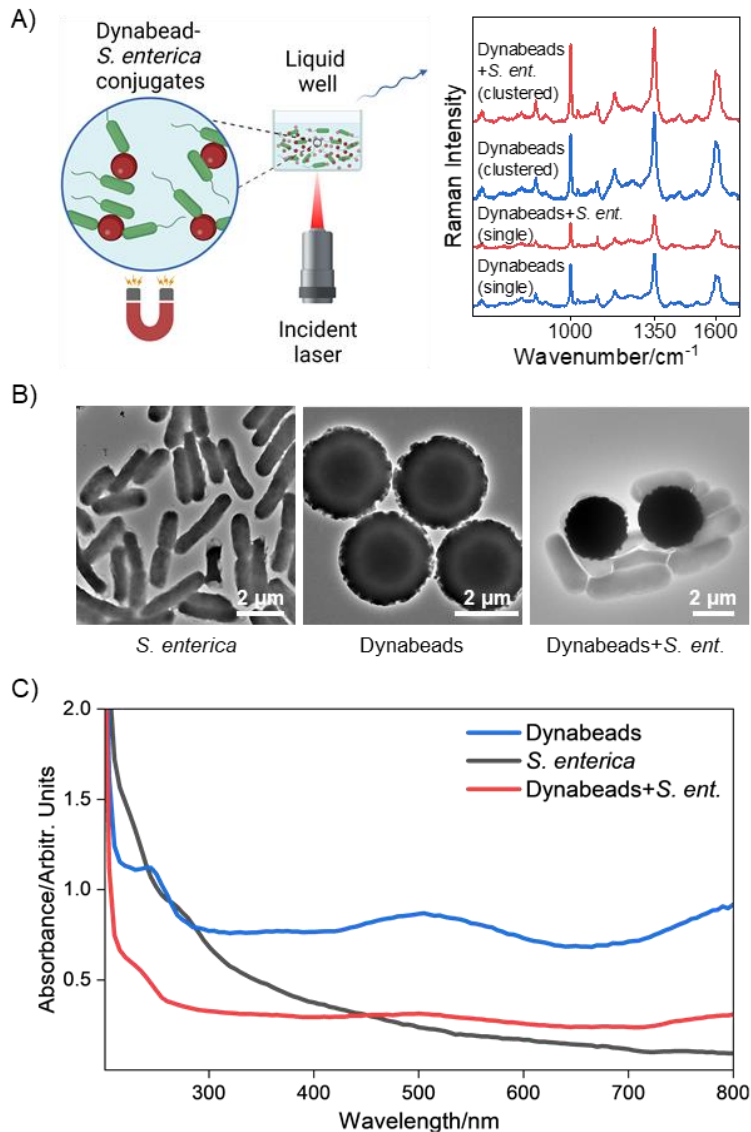

**Figure 1.** Overview of Raman setup and Dynabeads employed. (A) Liquid well imaging setup with *S. enterica* (green) and Dynabeads (red) suspended in DIW being pulled down to the imaging surface with a magnet prior to Raman interrogation with a 785 nm laser. Schematic created with Biorender.com (B) TEM of (left) rod-shaped *S. enterica* bacteria target, (middle) 2.5 μm Dynabeads, and (right) *S. enterica*-bound Dynabeads, showing multiple cells binding to each bead and across beads. (C) UV-Vis absorption spectra of Dynabeads, *S. enterica*, and *S. enterica*-bound Dynabeads, showing broad and large absorbance of Dynabeads primarily due to the iron oxide core as confirmed by later analyses.

As shown in Figure 2, Raman spectra of dried samples of single and clustered Dynabeads with and without conjugation to *S. enterica* show a unique signature from a single, 30 s acquisition at 10 mW laser power. This property, when combined with emerging techniques for the separation of target bound and unbound Dynabeads[10], could enable sensitive, specific detection of target molecules. Brightfield images in Figure 2A show the spots from which measurements in Figure

2B were collected. As expected, at this acquisition condition, *S. enterica* alone has a weak signal with low intensity, making it difficult to identify signature peaks typically[21] occurring around 1000, 1350, 1450, and 1660 cm$^{-1}$ (Figures 2B & S5). In a strong contrast, spectra from bead-containing samples display clearly defined peaks at high intensities with a unique Raman signature of Dynabeads highlighting their Raman reporter property (Figures 2B & S5). We attribute characteristic peaks from the Dynabeads at 1000 and 1600 cm$^{-1}$ to aliphatic and aromatic C-C stretching vibrations from the polystyrene coating of the beads (Figure S6). The major peak near 1350 cm$^{-1}$ is attributed to second order Raman scattering of the iron oxide core.[22,23] Contributions from antibody coating are negligible at these acquisition parameters (Figure S7 and S8).

Of note, these peaks are prominent both in single and clustered samples allowing detection of signals even from individual beads bound to targets. This is particularly important as target bacteria in fluids of interest such as wastewater, blood and similar biological fluids tend to be very low in number. Notably, higher intensity signature is observed with clustering of beads which is likely due to increased polystyrene and iron oxide content at the spot of optical interrogation, resulting in stronger Raman signals (Figure 2C). Signal intensity difference is calculated with respect to the signal intensity of the *S. enterica* only sample analyzed from four different locations. In the case of single beads, the Raman signal intensity from *S. enterica*-bound Dynabeads seems to be dampened compared to signal from single unbound Dynabead (Figure 2C) with an average of 1.5X larger intensity reading at the three major peak locations 1000, 1350 and 1600 cm$^{-1}$. This effect could be due to multiple bacteria covering the surface of the polystyrene coating and iron oxide core, scattering the light and reducing optical accessibility for Raman interrogation, an issue that could be amplified in dried sample preparation due to additional refractive index contrast. Interestingly, in the case of clusters (Figure 3B), signal intensity is minimally and preferentially affected with an average of 1.1X larger intensity at 1000 and 1350 cm$^{-1}$, and 1.1X dampening at the 1600 cm$^{-1}$ peak in signal from *S. enterica*-bound Dynabeads compared with their unbound cluster counterparts. This could potentially be because of the locations of cells, which could be arranged around the outer edge of the cluster, leaving a central bead-only portion optically accessible, hence not affecting the signal intensity as much (Figures 2B-C and S5). We also note signal dampening could also be due to variation in bead-to-bead polystyrene and iron oxide distribution. As shown in Figure 1 and 5, TEM and SEM images show non-uniform polystyrene coating. EDX mapping also shows non-uniform distribution of iron (inset of Figure 5B). We would like to emphasize that even though on average such differences are noted, they are within the error range for the four different locations studied making it statistically insignificant when considering large scale measurements in practical applications.

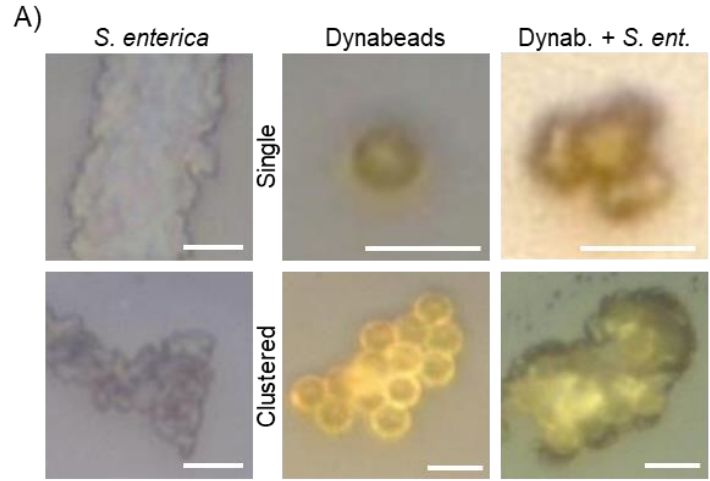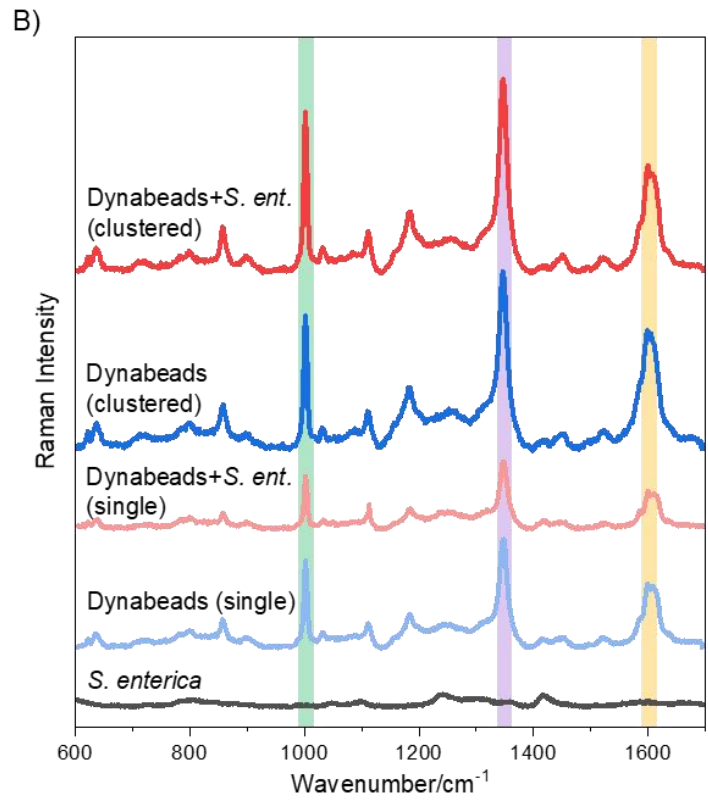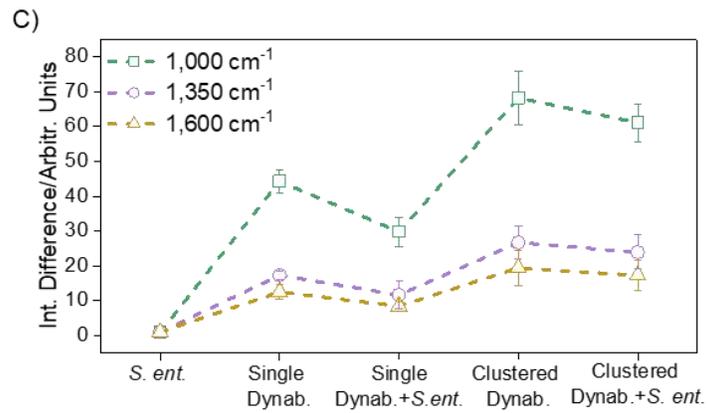

**Figure 2.** Raman interrogation of dried samples from single point detection with a 785 nm laser at 10 mW power for a single, 30 s acquisition. (A) Brightfield images of *S. enterica*, Dynabeads (single and clustered), and *S. enterica*-bound Dynabeads (single and clustered) showing areas where the incident laser was focused for Raman measurements. (B) Corresponding Raman spectra showing clear Raman signature of beads with major peaks at 1000, 1350, and 1600 cm$^{-1}$ which are primarily signatures of polystyrene coating (Figure S6) and iron oxide core. Minor contributions may be made from the anti-*Salmonella* antibody as shown in Figure S7, but antibodies are not a major source of Raman signal as found in Raman analysis of surface-activated, antibody-less beads (Figure S8). This unique Raman signature from Dynabead reporters is maintained upon conjugation with cells, demonstrating the signal is coming from the beads and highlighting their strong Raman reporter capability after target capture. (C) Analysis of Raman intensities from Dynabead samples with respect to Raman signature from bacteria alone. Intensities from selected wavenumbers (1000, 1350, and 1600 cm$^{-1}$) corresponding to the signature peaks of Dynabeads from each sample were divided with the respective intensities of *S. enterica* to calculate the intensity difference. All analysis was performed from data collected at four different sample locations, as shown in Figure S5. Overall, the highest intensity is recorded at the 1000 cm$^{-1}$ peak; single *S. enterica*-bound Dynabeads have lower signal intensities than single unbound Dynabeads, but clusters of bound beads have mostly comparable intensities at the signature peaks compared to clusters of unbound Dynabeads.

While strong, unique signatures in Figure 2 are achieved with a 30 s point acquisition using a 10 mW 785 nm laser intensity, these acquisition conditions may be time consuming for large scale applications. Thus,we demonstrate the high throughput target identification potential of Dynabead Raman reporters by capturing spectra from a large area of ~30 x 30 *μ*m  using only single shot 0.5 s acquisition with 7 mW laser power as shown in Figure 3 and Figure S9. Brightfield images and intensity maps show the intensity of the Raman shift coming from the site of Dynabeads, with maximum intensity observed at the center of bead clusters (Figure 3A). As shown, characteristic peaks from the beads are still identifiable (Figures 3B-C) even with markedly reduced detection times, demonstrating the potential for Dynabead Raman reporters in high-throughput single-shot wide-field Raman imaging systems. Bacteria alone have high signal count on the intensity map but have mainly background signal (intensity maps shown are prior to background subtraction) and show no distinct signature peaks at this acquisition time and laser power. The signal intensity from single Dynabeads in bound or unbound states is comparable whereas in the case of clusters some dampening is noted upon conjugation as described above.

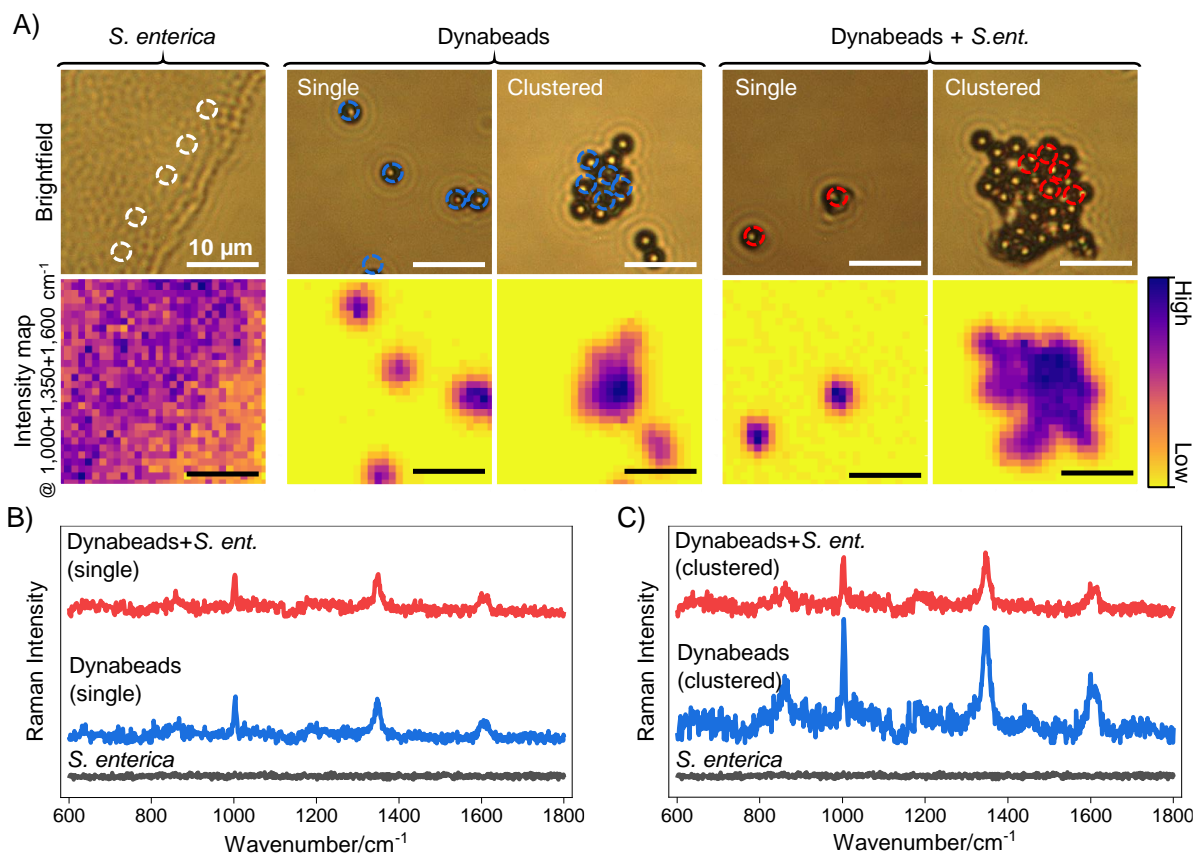

**Figure 3.** Raman mapping of dried samples at 0.5 s acquisition for high throughput signal detection. (A) Brightfield images and intensity maps of *S. enterica*, Dynabeads (single or clustered), and *S. enterica*-bound Dynabeads (single or clustered) with dotted circles showing regions where the incident laser was focused and whose intensity maps and spectra are shown below. All scale bars are 10 μm. (B) Raman spectra from *S. enterica* (black), single Dynabeads (blue), and single *S. enterica*-bound Dynabeads (red) collected with a single shot 0.5 s acquisitions with 7 mW laser power across a 30 x 30 μm area. Signature peaks of Dynabeads at 1000, 1350, and 1600 cm$^{-1}$ are noted. Spectra shows matching signatures in both bound and unbound beads, confirming that the signal source is the Dynabeads. (C) Raman spectra from *S. enterica* (black), clustered Dynabeads (blue), and clustered *S. enterica*-bound Dynabeads (red) with higher intensities observed for bead-only clusters.

We also demonstrate the application of Dynabeads in liquid format, which will enable both simultaneous isolation and detection in complex fluid samples directly from the source with high specificity when combined with emerging techniques such as density-shift separation of target bound and unbound Dynabeads, where Dynabead-target bacteria complexes can be separated from free Dynabeads in a density centrifugation step.[10] As shown in Figure 4, the Dynabeads' Raman reporter signature is preserved in liquid samples. Here, we interrogated solutions of bacteria and Dynabeads in DIW after magnetically concentrating *S. enterica*-bound Dynabeads to the bottom of the liquid well, which can be adopted as a scheme for identification of our target foodborne pathogen *S. enterica* from post-wash wastewater of fruits and vegetables. Note that, while the focus of this work was to demonstrate the Raman reporter property of Dynabeads, methods for the separation of target-bound Dynabeads from free Dynabeads such as those mentioned above will need to be implemented for specific detection of target molecules. Interestingly in liquid measurements, brightfield images before and after laser exposure show dislocation of *S. enterica* only and unbound Dynabeads-only samples upon laser exposure, resulting in small to undetectable signal from bead only samples despite their high intensity signals in dried samples (Figures 4A, S10, S11). This is due to unbound cells and beads freely moving through the liquid and dislocating after magnetic concentration and upon laser exposure due to heat-induced convective flow. Notably, however, both single and clustered *S. enterica*-bound Dynabeads remain stable post magnetic concentration and laser exposure and show strong reporter signature consistent with that of Dynabeads alone (Figures 4B-C, S10, and S11).

Though aggregation occurs even in Dynabeads alone, we believe conjugation to *S. enterica* results in strongly bound, higher mass clumps compared to the *Van der Waals* force-based clustering of unbound Dynabeads as single bacteria can bind to at least two beads at a time. Raman spectra from the selected regions in *S. enterica*-bound Dynabeads (dashed circles in Figure 4A) show similar patterns to those obtained from dried samples (Figure 2 and 3) with signature peaks near the wavenumbers of 1000, 1350, and 1600 $cm^{-1}$. As an additional advantage, in liquid samples, laser powers can be increased 10X that of dried format (75 mW vs 7 mW at 0.5 s acquisition) without melting of the Dynabeads as the water serves to dissipate the laser induced thermal effects. These results further emphasize that, through detection of Raman signal from Dynabeads, the rapid, dual isolation and indirect detection of target bacteria from liquid samples positive for bacteria contamination is possible and intrinsically leads to aggregates that concentrate to the imaging surface when an external magnetic field source is applied. We believe this assay is further enabled by and is complementary to emerging methods of separating target-bound and unbound beads[10] where the Raman signature from bacteria-bound beads serves as a rapid inline detection method even in cases where there are only a few bacteria in large volumes of solution, typical of target samples and routinely requiring culturing and staining steps after separation.

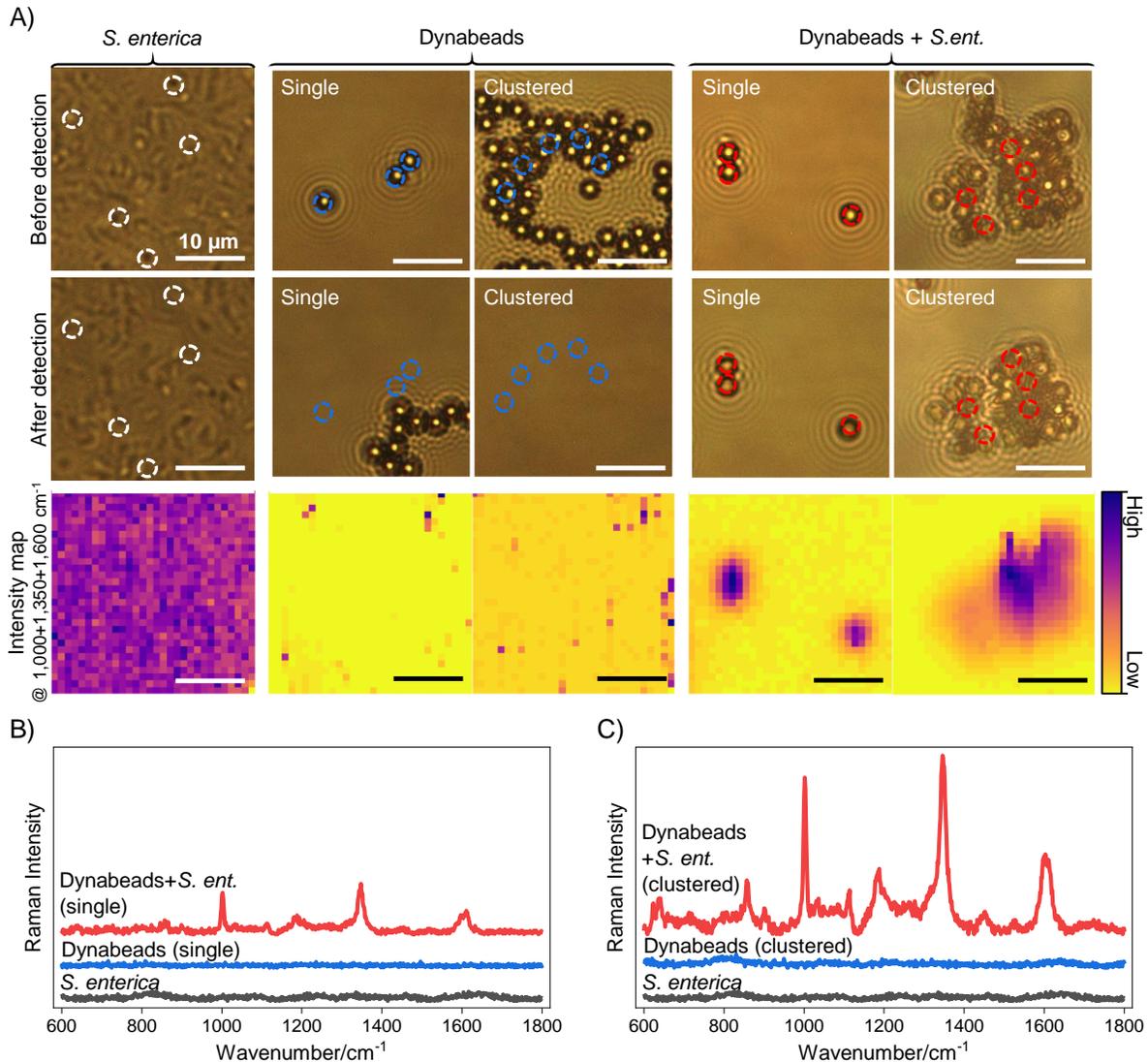

**Figure 4.** Raman interrogation in liquid samples with 0.5 s acquisition at 75 mW laser power and 785 nm laser. (A) Brightfield images and intensity maps of *S. enterica*, Dynabeads (single and clustered), and *S. enterica*-bound Dynabeads (single and clustered) showing dislocation of cell or bead only samples upon laser exposure. In contrast, cell-bound beads (single and clustered) stay in place after laser exposure. (B-C) Raman spectra of *S. enterica* (black) and single or clustered Dynabeads only (blue) and *S. enterica*-bound Dynabeads (red). Both single and clustered Dynabead only samples show inhibited detection due to dislocation of Dynabeads upon laser exposure from heat-induced convective fluid flow. *S. enterica*-bound Dynabeads, in both single and clustered format, remain in place at the bottom of the well upon laser exposure and show similar spectral features as dried samples, with larger intensity from bacteria-bound clusters.

We further characterize the origins of the Raman signature using surface morphology and material composition studies of the Dynabeads using scanning electron microscopy (SEM), transmission electron microscopy (TEM), and energy-dispersive X-ray (EDX) microanalysis as shown in Figure 5. As expected, Dynabeads have a relatively uniform size distribution and an uneven, porous surface morphology consisting of a polystyrene coating with a maximum thickness of ~150 nm and a ~2.44 μm iron oxide core confirmed by EDX analysis (Figure 5). A major carbon peak and minor copper and sulfur peaks are observed and attributed to the polystyrene coating, TEM grid, and Tosyl groups used to crosslink the antibodies to the bead surface respectively. As shown in the inset of Figure 5B, iron and oxygen are dispersed throughout the core of the bead. This is corroborated by prior studies, where Dynabeads were found to be composed of iron oxide nanoparticles with sizes ranging from 6-12 nm that occasionally form 20 nm clusters.[24] Focused-ion beam (FIB) imaging shown in Figure S12 did not show a distinct border between the inner core and the outer shell. Interestingly, while oxygen is uniformly distributed, iron shows a single-sided, preferential distribution consistently across all EDX analysis (Figures 5 & S13). This could possibly be due to part of the carbon-rich polystyrene internally coating some iron oxide core nanoparticles. A similar observation was reported in dextran-coated iron oxide nanoparticles.[25]

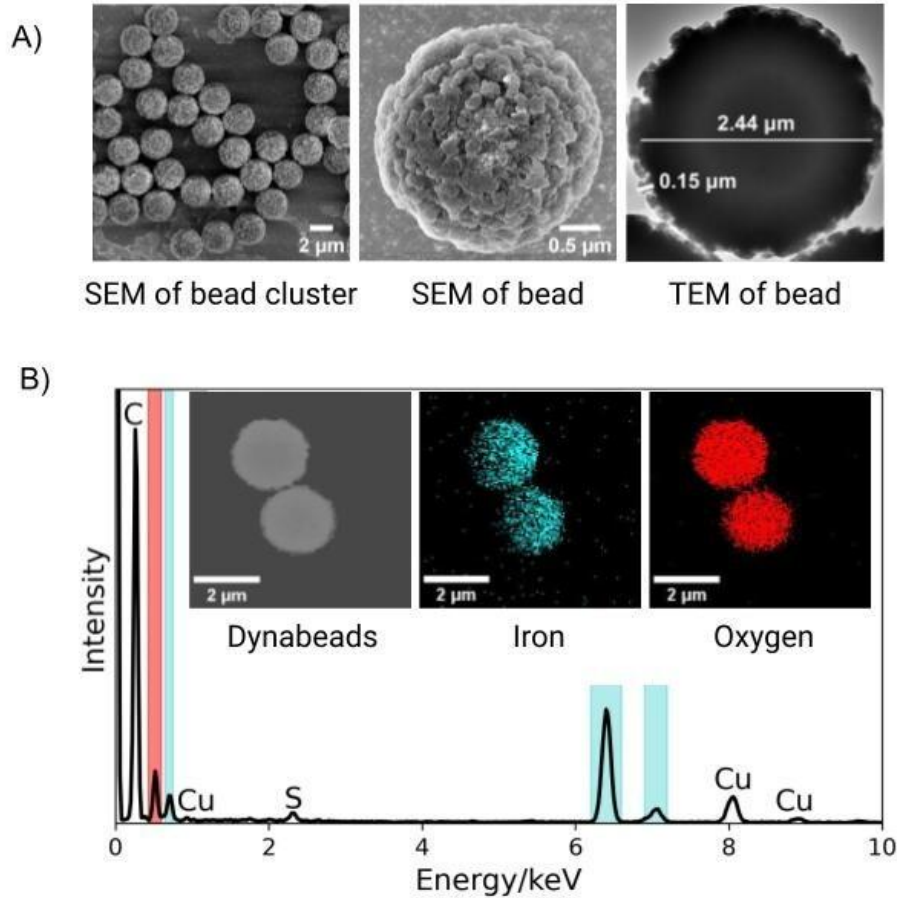

**Figure 5.** Characterization of Dynabeads anti-*Salmonella*. (A) SEM & TEM of Dynabeads showing (left) uniform size distribution, (middle) porous surface morphology, and (right) widths of iron oxide core (~2.44 μm) and polystyrene shell (~0.15 μm). (B) EDX spectrum of Dynabeads, showing significant contributions from iron (cyan) and oxygen (red). Inset shows EDX mapping image of two Dynabeads with corresponding regions of iron (middle) and oxygen (right). Oxygen distribution is uniform throughout the core, but iron distribution is right shifted. A major carbon peak and minor copper and sulfur peaks are observed and attributed to the polystyrene coating, TEM grid, and Tosyl groups respectively.

In summary, we have demonstrated the strong Raman reporter property of commercially available, versatile antibody-coated Dynabeads. These findings, with emerging methods for the separation of target-bound and unbound beads[10], enable the dual isolation and detection of targets via the recording of Dynabeads' Raman signature spectra. We note prominent spectral peaks originate from the polystyrene coating of the Dynabeads, as well as second order Raman scattering of the iron oxide core. We show that these beads can be deployed for the simultaneous capture and detection of biomolecules in liquid samples, dried samples and with rapid, single shot high throughput detection schemes. Of note, aggregation of *S. enterica*-bound Dynabeads dramatically increases Raman signal intensities compared to single bound beads due to higher polystyrene and iron oxide content at the location of measurement, but this effect can be dampened especially in dried samples due to bacteria covering the beads with additional refractive index contrast and scattering effect limiting access to the bead surface for optical interrogation. In addition, we have illustrated the use of Dynabeads' superparamagnetic property to magnetically concentrate *S. enterica*-bound Dynabeads to the imaging surface, resulting in stronger Raman signals in liquid samples. The theoretically unlimited target options and available customizations, besides the anti-*Salmonella* Dynabeads discussed here, expands capture and detection applications to diverse cell types and biomolecules. Particularly with advances in separating target-bound Dynabeads from unbound Dynabeads using a density media bi-layer[10], our demonstration opens a way for rapid interrogation of target-bound beads without additional culturing or staining steps. We believe our Raman-based detection approach sheds light on the unique Raman reporter property of versatile Dynabeads, overcoming current limitations in post-capture target detection without relying on unique plasmonic substrate engineering and lending itself to a wide variety of applications and workflows.

## Acknowledgements
This work was supported in part by the Koch Institute Support (core) Grant P30-CA1405, from MIT Laser Biomedical Research Center (NIH 5P41EB0158711), and from the National Cancer Institute and the Abdul Latif Jameel Water and Food Systems Lab (J-WAFS) at MIT. We would like to thank the Koch Institute's Robert A. Swanson (1969) Biotechnology Center for technical support in conducting TEM and EDX analysis, specifically Peterson (1957) Nanotechnology Materials Core Facility (RRID:SCR_018674). We would also like to thank Dr. Zhenyuan Zhang in the Electron Microscopy Lab at the Materials Research Laboratory at MIT for facilitating SEM and FIB imaging.

## Conflict of interest
JL, MM, NM, JWK, and LFT declare no conflict of interest. RK is co-inventor on a patent application related to isolation of cells using magnetic beads followed by their downstream detection and analysis using various methods including Raman spectroscopy.

## Author Contributions
**JL:** writing (original draft), writing (review and editing), conceptualization, investigation, visualization, formal analysis; **MM:** writing (original draft), writing (review and editing), conceptualization, investigation, visualization, formal analysis; **NM:** writing (original draft), investigation; **JWK:** writing (review and editing), resources, supervision, funding acquisition; **RK:** writing (review and editing), resources, supervision, funding acquisition; **LFT:** writing (original draft), writing (review and editing), resources, supervision, project administration, conceptualization

**ORCID**
*Jongwan Lee*: https://orcid.org/0000-0001-6736-3904
*Marissa McDonald*: https://orcid.org/0000-0003-4214-0303
*Rohit Karnik:* https://orcid.org/0000-0003-0588-9286
*Loza F. Tadesse*: https://orcid.org/0000-0003-2040-8145

## Methods

### *S. enterica* preparation and culture

*Salmonella enterica* (*S. enterica*, BAA-710™, ATCC, USA) was grown overnight in tryptic soy broth (TSB) purchased from Sigma-Aldrich (43592). Morning culture was done by reculturing 20 µl of overnight culture in 3 mL of fresh TSB for 3 hrs. The cells were incubated at 37 °C, shaken at 400 RPM with 0.1% $CO_2$. Subsequently, the cells were washed using phosphate buffer saline (PBS, 21-040, Corning®) via centrifugation for 10 mins at 3000 relative centrifugal field (RCF). Quantification of the cells was done using a disposable hemocytometer (inCYTO C-Chip™, DHC-S02, SKC Inc., Korea).

### Conjugation/binding of anti-*Salmonella* Dynabeads with *S. enterica*

Anti-*Salmonella* Dynabeads (71002, ThermoFisher Scientific Inc., USA) and *S. enterica* with concentration ratio of 1:0.5 were incubated in bovine serum albumin (BSA, A3294, Sigma-Adrich®, USA) -blocked 2 mL eppendorf tubes. The total volume of the reaction was 1 mL in a mixture of PBS with 0.05% Tween® 20 (PBST, P1379, Sigma-Adrich®, USA). The reactants were first vortexed for 1 min then swirled at ambient temperature for 20 min.

### Sample preparation for Raman collection

For Dynabead-containing samples (Dynabeads alone and *S. enterica*-bound Dynabeads), a magnet was placed near the sample-containing tube, forming a sample pellet on the tube wall. The buffer was pipetted out and replaced with fresh DIW. For *S. enterica*, the cell-containing tube was centrifuged for 10 min at 3,000 RCF and the supernatant buffer was replaced with DIW. As shown in Figure S4, each sample was washed twice with DIW to remove excess buffer for Raman spectral interrogation. To obtain Raman measurements in liquid, 500 $\mu$L of liquid from each sample was injected into the silicone isolator (hole with 13 mm-diameter and 2.5 mm-depth, 665307, Grace Bio-Labs, USA)-attached quartz coverslip (25.4 × 25.4 × 0.2 mm$^3$, 1×1×.2, Technical Glass Products, Inc., USA), and a magnet was placed underneath the well for 10 min to pull cell-bound Dynabeads down to the quartz coverslip. The magnet was then removed, and the well was placed on the Raman spectroscopy system for spectra collection. For the interrogation of dried samples, 2.5 $\mu$L droplets of each sample were drop casted on a quartz coverslip. Nitrogen gas with a pressure of 10 kPa was then blown for 10 min above the deposited droplets to accelerate drying. The samples were then placed on the Raman spectroscopy system for spectra collection. We confirm that the chemicals in the original stock buffers for cells and Dynabeads–TSB for *S. enterica* and PBS, BSA, and sodium azide (NaN$_3$, S2002, Sigma-Aldrich®, USA) for Dynabeads–do not have signature peaks in their Raman spectra, as shown in Figure S14. To further investigate the origin of signature peaks of Dynabeads anti-*Salmonella*, Dynabeads lacking surface antibodies but maintaining surface activation groups and polystyrene coating (14203, Dynabeads M-280 Tosylactivated, ThermoFisher Scientific Inc., USA) were investigated (Figure S8). Note that while the exact surface groups used to manufacture

Dynabeads anti-*Salmonella* are deemed proprietary information, tosylactivated beads are the only recommended beads for bacterial targets, and in addition, the particular Dynabeads M-280 Tosylactivated beads also match the size of Dynabeads anti-*Salmonella*. For this reason, these beads were considered to be the closest match to Dynabeads anti-*Salmonella* minus the surface antibodies. These beads were washed and drop cast onto the quartz substrate for Raman data collection, similar to the prior sample preparation protocol for Dynabeads anti-*Salmonella*. A polyclonal goat anti-*Salmonella* CSA-1 antibody powder (5310-0322, The Native Antigen Company, UK) was dissolved in DIW and then prepared for Raman spectra collection after casting and drying it following the same aforementioned procedure (Figure S7). Polystyrene (PS) powders with different molecular weights were purchased from Sigma-Alrich® (35000 (PN: 331651) and 350000 (PN:441147) g/mol, and were directly measured in their Raman spectra as their original state.

**Raman spectra collection**

We utilized two Raman spectroscopy systems for collecting Raman spectra from samples with different detection modes; single point detection and mapping mode. For single point detection, the InVia$^{TM}$ Reflex Raman system (Renishaw plc., UK) was utilized to detect Raman spectra from a 1 μm-sized exposing laser spot on the dried sample (*i.e.*, static detection). Single, 30 s acquisitions were conducted to measure spectra from Dynabeads anti-*Salmonella* (P = 10 mW), *S. enterica* (P = 10 mW), anti-*Salmonella* antibody (CSA-1, P=10 mW), and Dynabeads M-280 Tosylactivated (P = 5 mW) in Figures 2, S7, and S8. For higher resolution spectra of biologics, 1 min acquisitions at 100 mW power was used for Raman interrogation of *S. enterica* and anti-*Salmonella* antibody (CSA-1) in Figures S5 and S7. A customized inverted Raman system was utilized to detect Raman spectra from both high-throughput dried and liquid samples and visualize their intensities with single, 0.5 s acquisitions at 7 or 75 mW power from a field of view (FOV) of 30×30 $\mu m^2$ (*i.e.*, mapping).[26] Each FOV was divided into 900 pixels (30×30 pixels, 1×1 $\mu m^2$/pixel). We detail the experimental conditions for each Raman spectroscopy measurement in Table S1. We applied the magnet prior to Raman interrogation, but the exact methodology for magnetic concentration is adjustable to fit other applications and Raman systems, including sustained magnetic field application. After collecting Raman spectra data, we performed polynomial fitting-assisted background-subtraction using Lieberfit.[27] For drawing intensity maps, we picked wavenumbers corresponding to the signature peaks of Dynabeads; 1000, 1350, and 1600 cm$^{-1}$, and combined their intensities for clear differentiation of samples in the intensity map. All data, including Raman spectra from single point detection and mapping, were plotted using OriginPro software (OriginPro 2023, v10.0.0.154, OriginLab Corp., USA).

**TEM & EDX analysis**

For imaging under transmission electron microscopy (TEM), 10 μL of sample and buffer-containing solution was dropped on a 200-mesh copper grid (Electron Microscopy Sciences, USA) coated with a continuous carbon film and dried at room temperature. The grid was mounted on a JEOL single tilt holder equipped in the TEM column. Imaging on a JEOL 2100 FEG microscope was done using the largest area size of parallel illumination beam and a condenser aperture 100 μm in diameter. The microscope was operated at 200 kV with a magnification in the ranges of 3,000 to 600,000 for assessing particle shape, size, and atomic arrangement. All images were recorded on a Gatan Side mounted camera and shown in Figure 1. STEM imaging was done by a HAADF (high-angle annular dark field) detector with 0.5 nm probe size and 12 cm camera length. X-Max 80mm$^2$ EDX (Oxford Instrument, UK) was used for chemical information mapping of samples (Figure 5).

**SEM & FIB imaging**
To prepare samples for scanning electron microscopy (SEM) and focused ion beam (FIB) analysis, silicon wafers were cut and cleaned with isopropyl alcohol and acetone before mounting on a stub. The stub surface was then sputtered with a 30 nm gold coating. 20 μL of Dynabeads anti-*Salmonella* suspended in DIW at a concentration of $10^7$ beads/mL were drop casted onto the wafer and dried with nitrogen gas at a pressure of 10 kPa before an additional layer was added. Following drying of both layers, the sample area was coated with another 10 nm of gold to reduce charging effects under the electron beam. The beads were then imaged at a working distance of 4.0 mm and magnification ranging from 4989-35000 with the FEI Helios Nanolab 600 Dual Beam System at 5 kV and a current of 86 pA. The horizontal field width ranged from 3.66-25.7 μm. 7 μm-deep cross-sectional cuts were made with FIB analysis at 30 kV and a current of 0.46 nA. SEM images were then taken at a stage tilt of 52° and 8012 magnification with a horizontal field width of 16 μm; these images are shown in Figure S13.

**Fluorescence microscopy of anti-*Salmonella* Dynabeads and *S. enterica***
Each 100 μL sample of cells alone, beads alone, or *S. enterica*-bound beads in PBS buffer was stained with 1 $\mu$L of SYTO$^{TM}$ 9 (S34854, ThermoFisher Scientific Inc., USA) solution in DMSO ($C$= 5mM) at 4 °C overnight and observed using a CCD camera (Andor iXon, Oxford Instruments, UK)-equipped fluorescence microscope (Nikon Eclipse TE200U, Nikon Instruments, NY) after buffer change with fresh PBS.

**UV-Vis spectroscopy**
For UV-Vis analysis, *S. enterica*-bound Dynabeads were washed by magnetization using PBST to remove unbound *S. enterica.* An Agilent technologies Cary 60 UV-Vis was used to study the optical properties of the *S. enterica*-bound Dynabeads, anti-*Salmonella* Dynabeads alone, and *S. enterica* alone.

# Supplementary Information

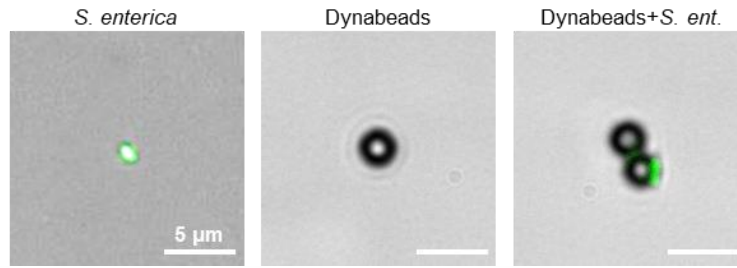

**Figure S1.** Fluorescence microscopy of *S. enterica* (left), Dynabeads (middle), and *S. enterica*-bound Dynabeads (right) showing size, shape, and tight binding interaction. All scale bars are 5 µm.

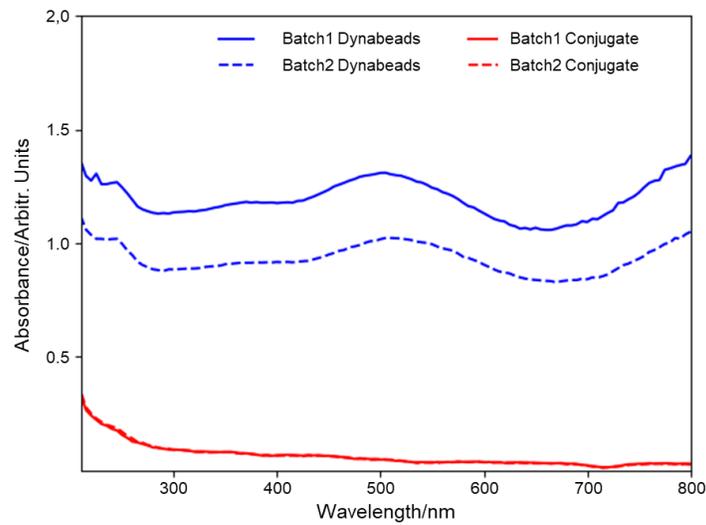

**Figure S2.** UV-Vis spectra of two batches of Dynabeads and *S. enterica*-bound Dynabeads showing consistent absorbance trends across batches. Decreases in absorbance from Batch 2 could be due to the degradation of Dynabeads' as they decay past their shelf life.

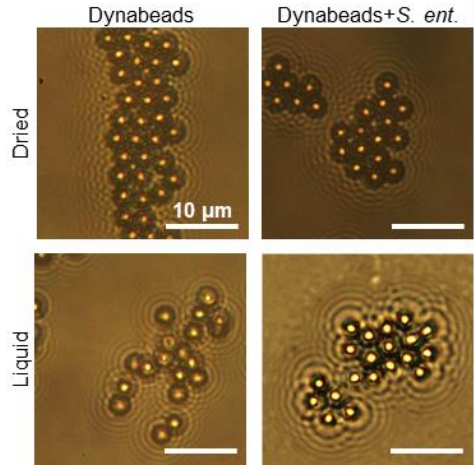

**Figure S3.** Clustering of Dynabeads in dried and liquid samples before and after conjugation to *S. enterica*. All scale bars are 10 μm.

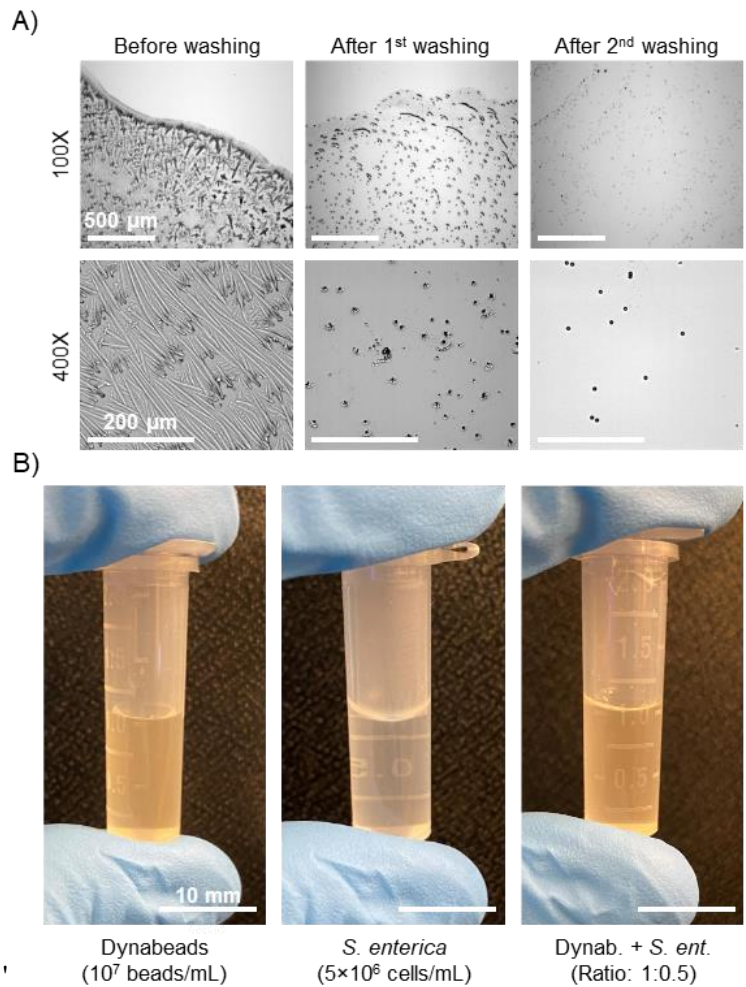

**Figure S4.** Preparation of samples for Raman interrogation. (A) Optical images of dried Dynabeads with 0, 1, and 2 washing cycles. Most residues in the stock buffer (sodium azide and BSA) can be removed after two cycles of washing. (B) Dynabeads and *S. enterica* samples were prepared at a fixed concentration of $10^7$ beads/mL and $5\times10^6$ cells/mL respectively. For conjugation, a ratio of 1:0.5 Dynabeads to *S. enterica* was used.

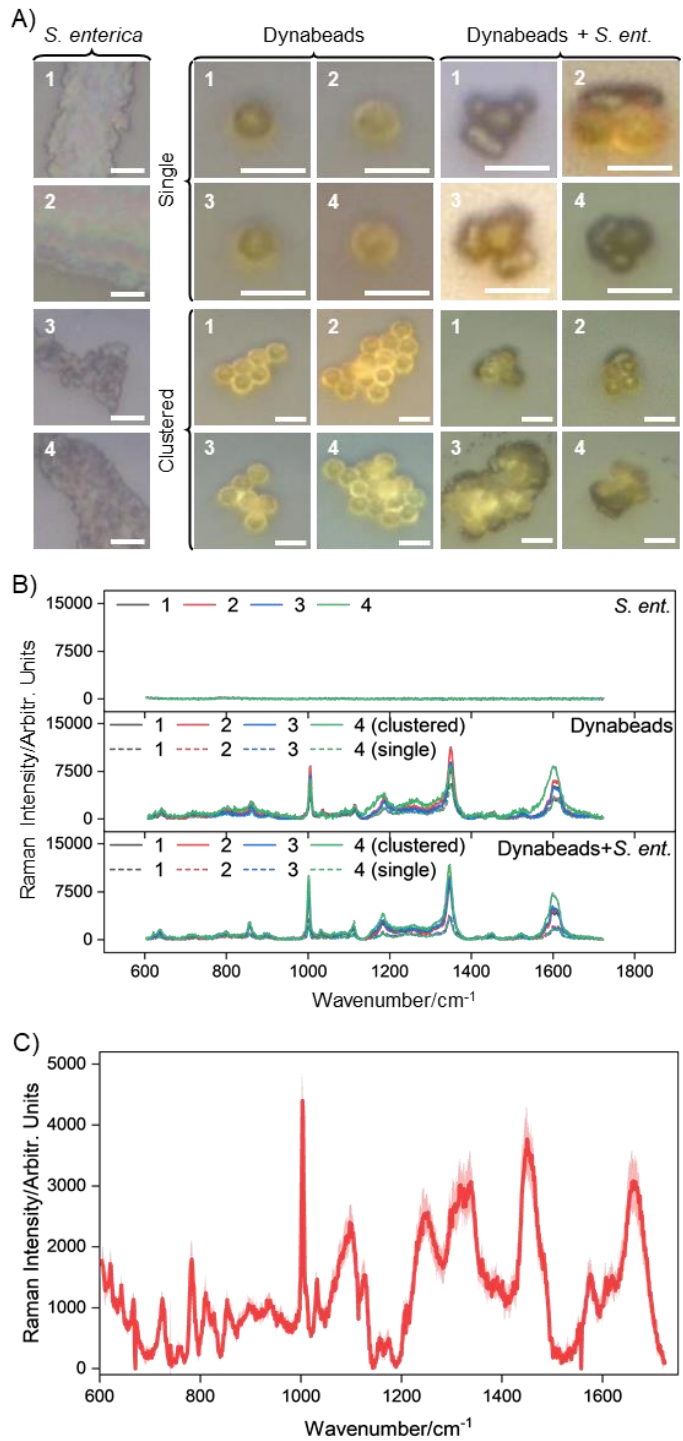

**Figure S5.** Raman spectra from dried samples. (A) Brightfield images of the four different sample locations where spectral data was collected (scale bar = 5 µm). (B) Raman spectra from *S. enterica* (top), single- and clustered Dynabeads (middle), and single- and clustered *S. enterica*-bound Dynabeads. (C) Raman spectra of *S. enterica* with modified acquisition settings for a higher resolution ($C_{S.\ ent}$= 3×10$^9$/mL, laser power= 100 mW, acquisition= 60 s). Signature peaks can be seen at 1000, 1350, 1450, and 1660 cm$^{-1}$, similar to previous literature reports.

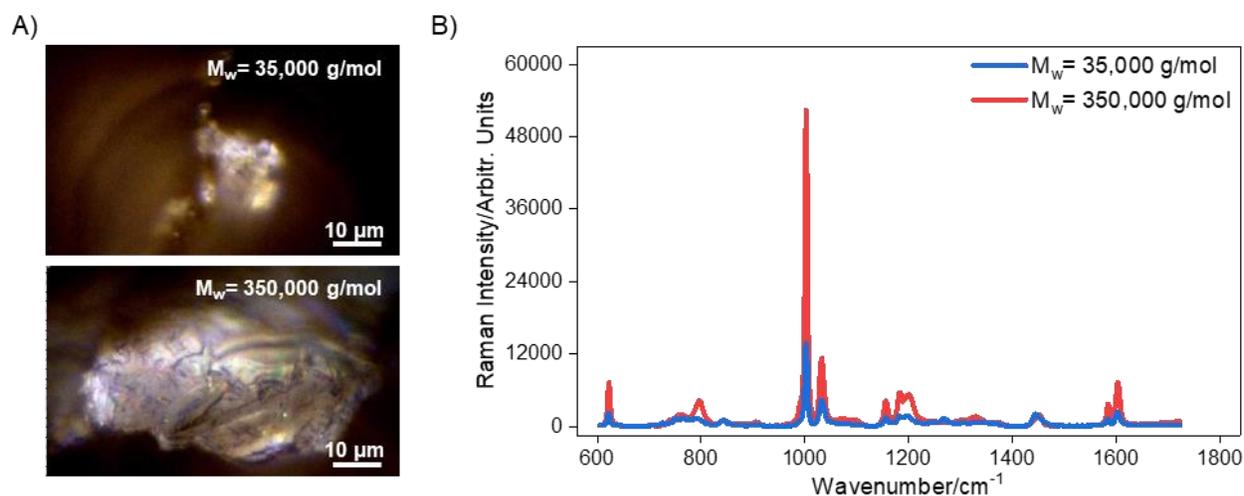

**Figure S6.** Raman spectra from polystyrene (PS). (A) Brightfield images of PS with different molecular weights ($M_w$): 35,000 and 350,000 g/mol. Images show where the incident laser was focused for Raman interrogation. (B) Raman spectra of PS of different $M_w$. Prominent peaks at 1000 and 1600 match observation in Dynabeads-containing samples, indicating large contributions of polystyrene to Raman spectra of Dynabeads.

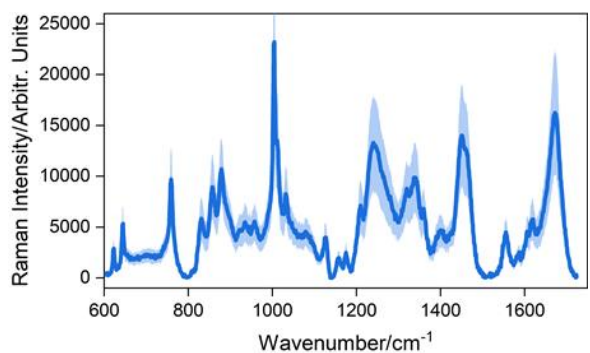

**Figure S7.** Raman spectra of anti-*Salmonella* antibodies with 1 min acquisition at 100 mW. Major peaks appear at 1004, 1243, 1452, and 1674 cm$^{-1}$.

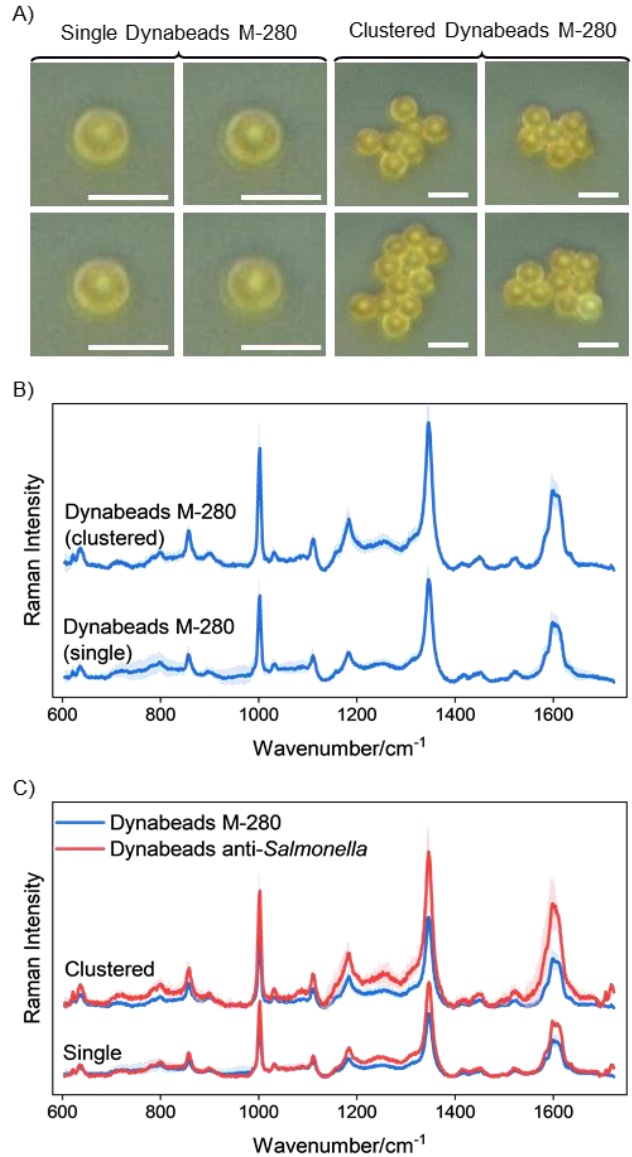

**Figure S8.** Raman spectra of Dynabeads M-280 Tosylactivated, which are surface-activated Dynabeads lacking surface antibodies. (A) Brightfield images of clustered and single Dynabeads M-280 Tosylactivated (Scale bar: 5 µm). (B) Raman spectra of single and clustered Dynabeads M-280 Tosylactivated showing strong peaks around 1000, 1350, and 1600 cm$^{-1}$. (C) Overlap of Raman spectra from Dynbeads M-280 Tosylactivated and Dynabeads anti-*Salmonella*, showing a direct match of Raman signal and indicating little to no contribution from surface antibodies on Dynabeads anti-*Salmonella*.

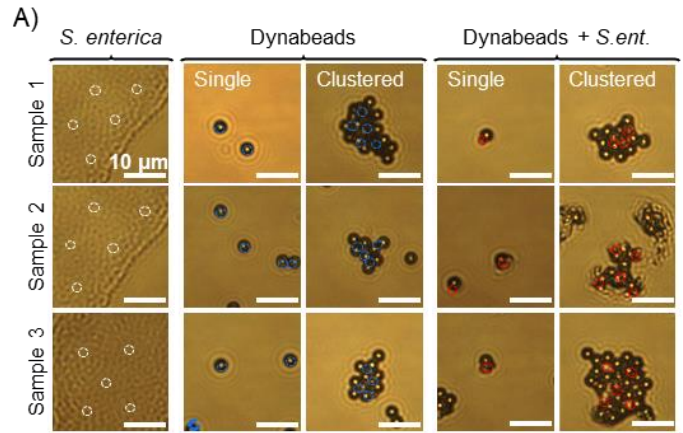
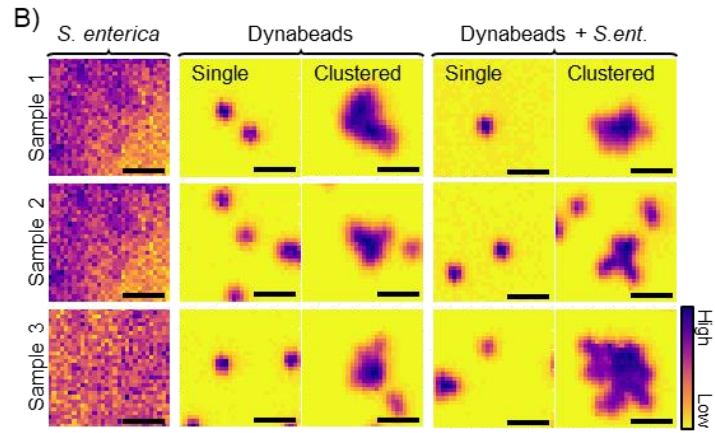
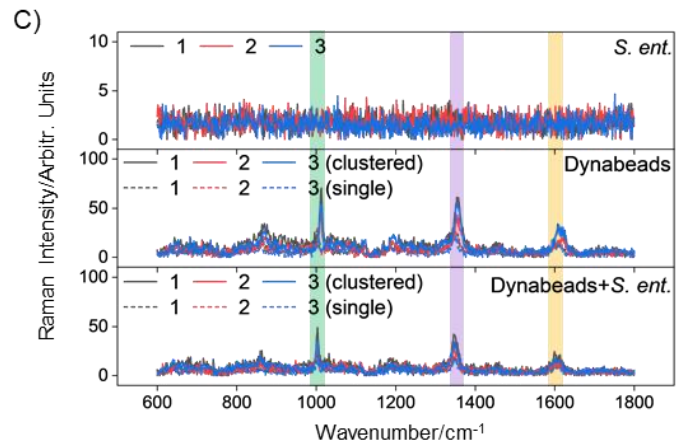
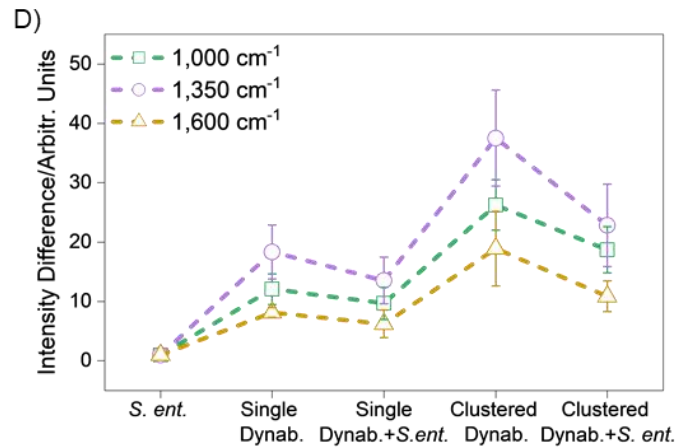

**Figure S9.** Raman map from dried samples with 785 nm laser under 0.5 s exposure and 7.5 mW power. (A) Data from three different locations were collected and analyzed (scale bar = 10 μm). Highly concentrated pellets were selected as *S. enterica* region. Locations with visible *S. enterica* and Dynabeads in the form of single and clustered formats, are selected as *S. enterica*-bound Dynabeads region. (B) Intensity maps after combining intensities at selected wavenumbers; 1,000, 1,350, and 1,600 cm$^{-1}$. At this acquisition parameter no detectable signature is observed from *S. enterica* only sample, mainly background intensity is detected. (C) Raman spectra from *S. enterica* don't show specific signature peaks under the acquisition parameters. In contrast, Dynabeads alone and *S. enterica*-bound Dynabeads show strong signature peaks from locations circled in dotted lines in the bright field images in A with notable peaks at 1000, 1350, and 1600 cm$^{-1}$. (D) Intensity difference factors as compared to the bacteria only baseline shows a similar trend as the single point measurement in Figure 2 with overall higher intensity recorded from clustered Dynabeads.

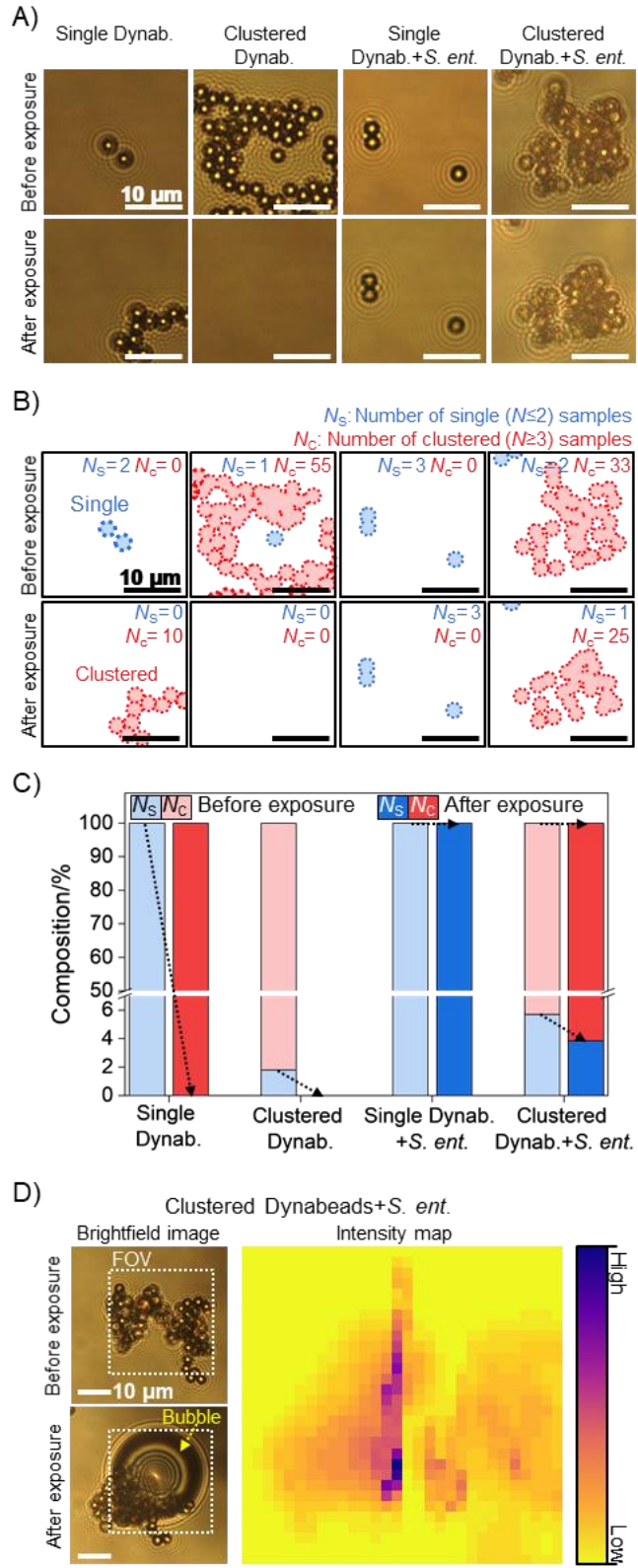

**Figure S10.** Dislocation of Dynabeads and *S. enterica*-bound Dynabeads before/after exposure to 75 mW-powered laser (*λ*= 785 nm) for 0.5 s. (A) Each sample was observed using a brightfield microscope, top showing before and bottom after exposure to laser. (B) Classification of samples in FOV, as single (blue circles, $N\leq2$) and clustered (red circles, $N\geq3$) forms, showing distinct differences in Dynabeads and *S. enterica*-bound Dynabeads; all of Dynabeads were dislocated, but none- to few *S. enterica*-bound Dynabeads did. (C) Quantitative analysis on the composition of the number of single ($N_S$) and clustered ($N_C$) samples. For Dynabeads, all samples were entirely dislocated; two single Dynabeads were observed in FOV for single bead measurement before laser exposure but they were dislocated and replaced by a cluster after exposure. In the case of single bead measurement, pre-exposure, one single Dynabead (1.2 % composition) and a cluster of 55 Dynabeads (98.2 %) were observed. Post-exposure, all of the Dynabeads were dislocated from FOV, resulting in 0 % of each composition. For single, *S. enterica*-bound Dynabeads, three single samples (100 % in the composition of the singles) were observed, and they retained their position post laser exposure. For clustered *S. enterica*-bound Dynabeads, two singles (5.7 % in composition)- and cluster of 33 (94.3 %) were observed pre-exposure, and one single (3.8 %)- and cluster of 25 (96.2 %) were observed post- exposure. (D) Generation of bubbles inside clustered *S. enterica*-bound Dynabeads induced by the laser exposure.

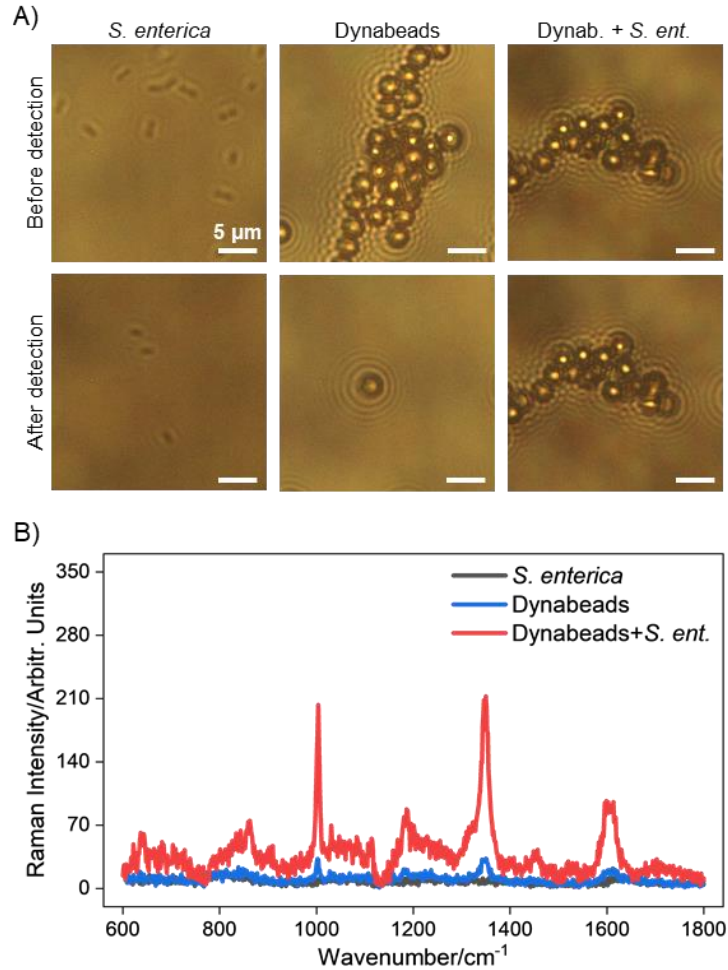

**Figure S11.** Effects of laser exposure on sample dislocation in liquid. (A) Brightfield images of each sample before- and after laser exposure during Raman interrogation. Only *S. enterica*-bound Dynabeads remain in place post-exposure. (B) Raman spectra from each sample at 75 mW-powered laser ($\lambda$= 785 nm) for 0.5 s. The spectra from *S. enterica*-bound Dynabeads showed the highest intensities compared to other samples.

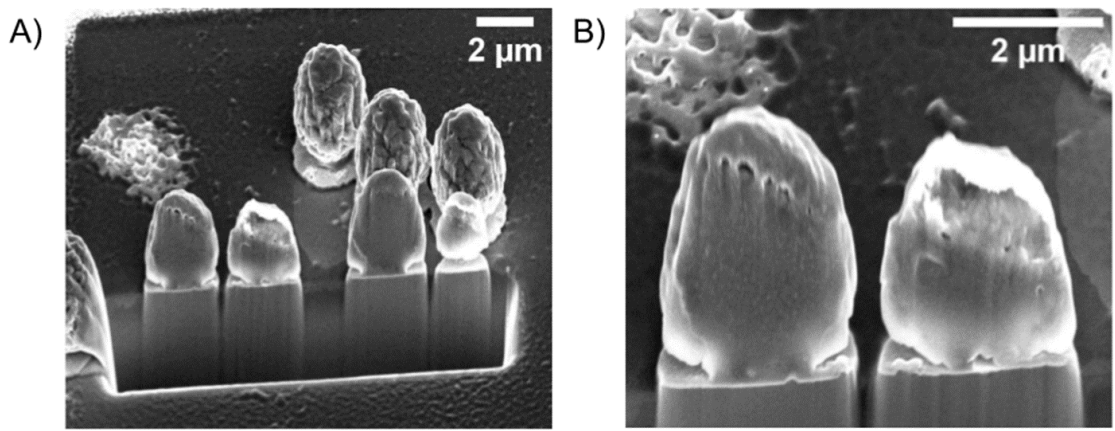

**Figure S12.** FIB analysis of Dynabeads. (A) Cross sectional image of 4 beads compared to 3 uncut beads also in the field of view. (B) Close up image of cross sectional view of Dynabeads exposes amorphous structure with no specific physical pattern of iron oxide core particles except for faint white dots.

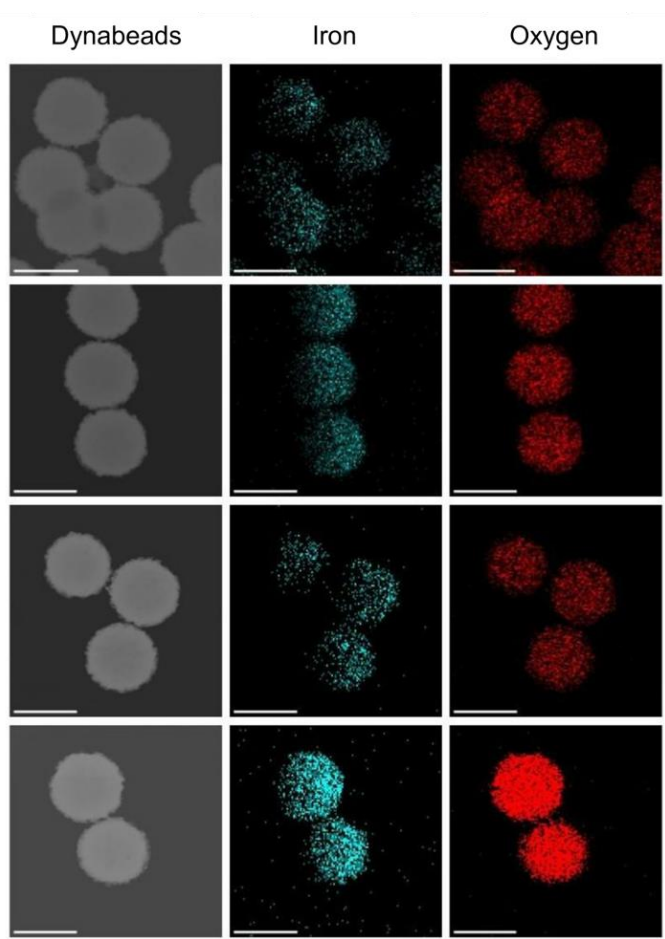

**Figure S13.** EDX images of Dynabeads (left) with corresponding regions of iron (middle) and oxygen (right). Oxygen distribution is uniform throughout the core, but iron distribution favors one side of the bead. All scale bars are 2 μm.

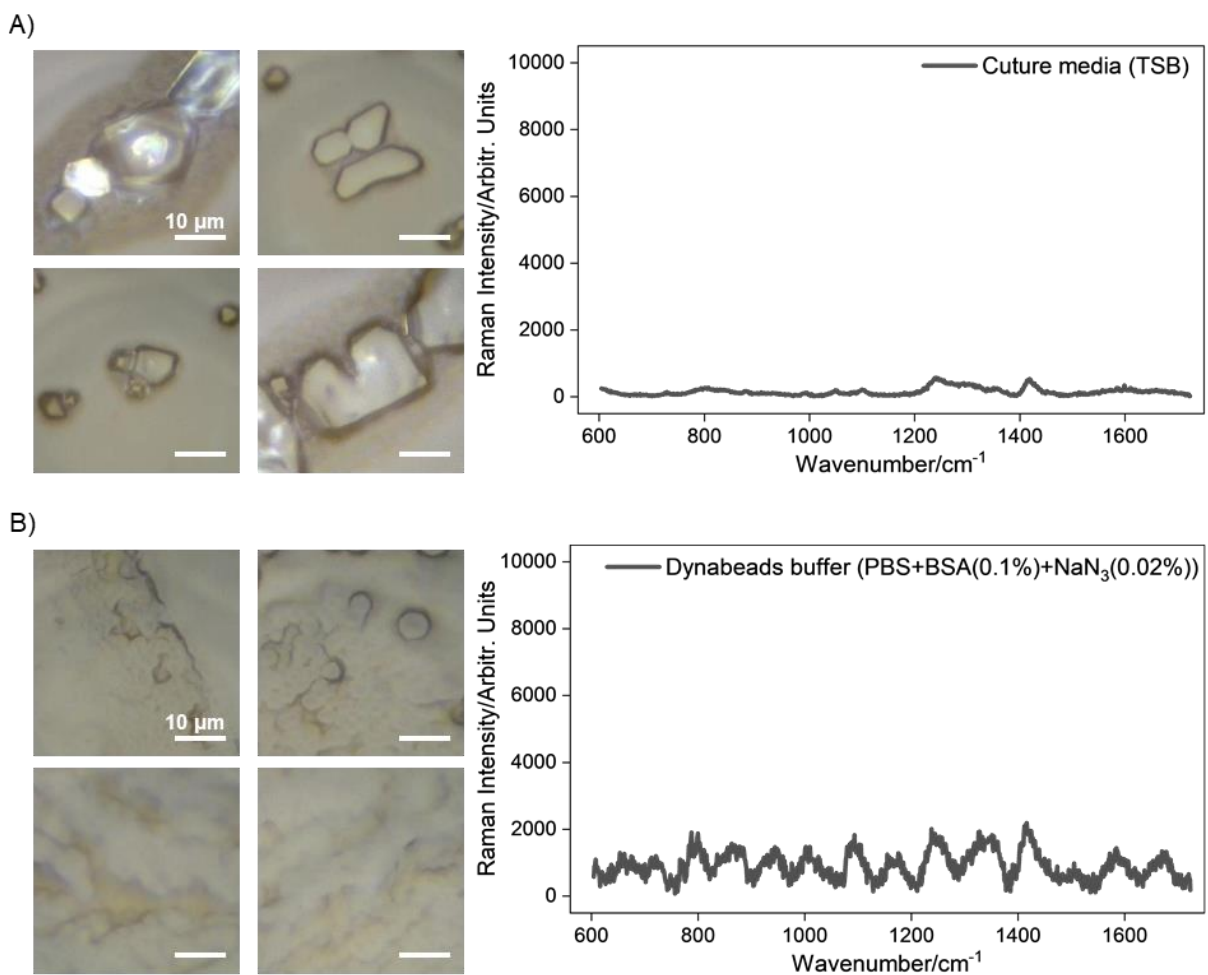

**Figure S14.** Raman interrogation of buffer solutions. (A) Brightfield images and Raman spectra from dried cell culture media (TSB). (B) Brightfield images and Raman spectra from Dynabead stock buffer (PBS, BSA, sodium azide). Both A and B demonstrate that the observed Raman signal of Dynabeads is not coming from residual buffer solution.

**Table S1.** Experimental conditions of Raman spectroscopy systems used in this study.

| System | Detection method | Target sample | Sample condition | Laser configuration | | | | |
|---|---|---|---|---|---|---|---|---|
| | | | | $\lambda$/nm | Spot size /$\mu$m | Power /mW | Exposure time/s | Accumulation |
| Renishaw Invia Reflex | Single point | Dynabeads anti-*Salmonella*, *S. enterica*, anti-*Salmonella* antibody (CSA-1), polystyrene ($M_w$= 35000, 350000 g/mol), Dynabeads buffer, TSB | Dried | 785 | 1 | 10 | 1 | 30 |
| | | Dynabeads M-280 | | 785 | 1 | 5 | 1 | 30 |
| | | Anti-*Salmonella* antibody (CSA-1), | | 785 | 1 | 100 | 2 | 30 |
| Customized | Imaging mode (30×30 $\mu$m) | Dynabeads anti-*Salmonella*, *S. enterica* | Dried | 785 | 1 | 7 | 0.5 | 1 |
| | | | Liquid | 785 | 1 | 75 | 0.5 | 1 |